\documentclass[%
prl,
 amsmath,amssymb,
 reprint,%
]{revtex4-2}
\usepackage{graphicx}% Include figure files
\usepackage{dcolumn}% Align table columns on decimal point
\usepackage{bm}% bold math
\usepackage{color}
\usepackage{float}
\usepackage[utf8]{inputenc}
\usepackage[T1]{fontenc}
\usepackage{mathptmx}
\usepackage{subfig}
\usepackage[inline]{enumitem}
\usepackage{caption}
\newcommand{\bogus}[1]{{}}
\begin{document}

\title{Collisionless ablative plasma shocks}

\author{Yanzeng Zhang and Xian-Zhu Tang}
\affiliation{Theoretical
  Division, Los Alamos National Laboratory, Los Alamos, New Mexico
  87545, USA}

\begin{abstract}
  An ablative plasma shock can emanate from the interface between a
  cold/dense plasma and a hot/dilute ambient plasma, where the plasma
  mean-free-path is much longer than the temperature gradient
  length. The shock is driven by thermal flux from the hot plasma into
  the cold plasma, primarily through tail electrons mediated by an
  ambipolar electric field, and it propagates into the ambient
  hot/dilute plasma.  Since the collisional mean-free-path is usually
  much longer than the Debye length, the ablative plasma shock is
  mostly collisionless, with the shock front width set by the upstream
  hot plasma Debye length and the shock speed by the downstream cold
  plasma sound speed. The shock heating of ions is extremely efficient
  via collisionless mixing of upstream hot ions and downstream cold
  ions, both of which have been converted into shock-front-bound flows
  accelerated by the ambipolar electric field that has a deep potential
  well anchored inside the shock front.
\end{abstract}

\maketitle

Explosions in the atmosphere provide a canonical example of shock
formation in neutral gases.  The shock width is known to be set by the
collisional mean-free-path $\lambda_\textrm{mfp}$ of the neutral
gases.  As a result, gas shock has a much wider front in rarefied
gases at high altitude compared with that in the low-altitude
atmosphere.  In an unmagnetized plasma or equivalently in a low-beta
plasma but strictly along the direction of the magnetic field, there
is another characteristic length that can be much shorter than the
plasma mean-free-path $\lambda_\textrm{mfp}.$ That is the Debye length
$\lambda_D$ associated with local charge separation in a plasma. The
analogous high-altitude atmospheric explosion in a plasma can drive a
similar shock but the shock width is set by $\lambda_D.$ A
collisionless plasma shock, which is electrostatic in nature and has
no counter-part in neutral gases, is realized in the asymptotic limit
of $\lambda_\textrm{mfp}/\lambda_D \gg 1,$ which can be easily
satisfied in a broad range of space, astrophysical, and laboratory
plasmas.  Much effort has been invested in this area of plasma shock
research through experiments, theory, and simulations since the 1960s,
and their findings can be found in the reviews of
Refs.~\cite{tidman1971shock,biskamp1973collisionless,eselevich1982shock,ryutovl2018collisional,sakawa2016collisionless}.

\begin{figure}[tbh]
\subfloat[]{\includegraphics[width=0.35\textwidth]{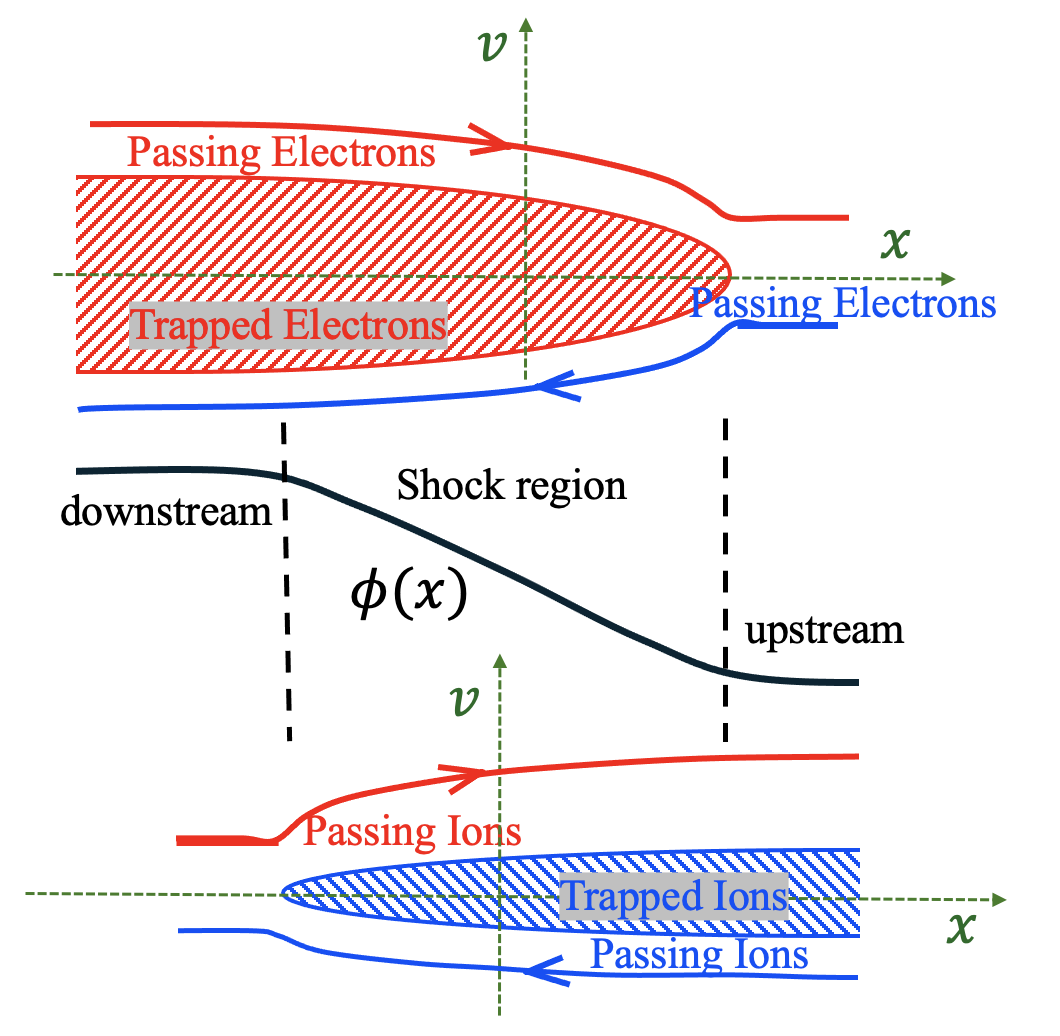}}
\hfill
\subfloat[]{\includegraphics[width=0.35\textwidth]{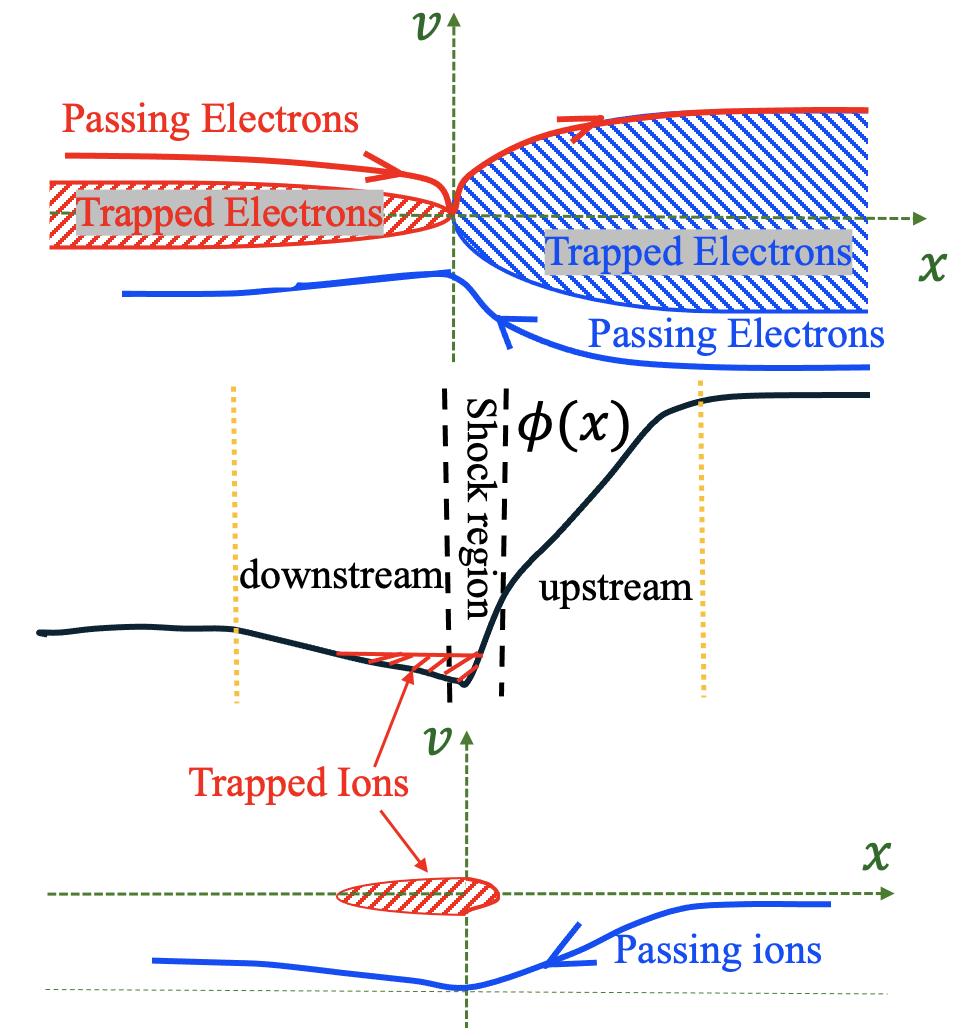}}
\caption{Schematic view of the explosive (a) and ablative (b) plasma
  shocks. In each diagram, the electrostatic potential is sketched in
  the middle, and the electron and ion distribution functions are
  shown above and below the electrostatic potential sketch,
  respectively. The particles coming from downstream (upstream) to
  upstream (downstream) are characterized by the red (blue) curves,
  where the trapped particles are shaded correspondingly. The vertical
  dashed black lines delimit the shock region, while in (b) we also
  label the recession fronts by the vertical dotted yellow lines that
  primarily limit the electrostatic potential variation.  }
\label{fig:diagram}
\end{figure}

A more recent surge of interest in plasma shock research is driven by
the study of plasma interactions with high-power
lasers~\cite{romagnani2008observation,koopman1967possible,haberberger2012collisionless,bell1988collisionless,wei2004ion,silva2004proton,nilson2009generation,morita2010collisionless,fiuza2012laser,fiuza2013ion,zhang2017collisionless}. The
rapid laser power deposition can induce localized over-pressured
high-density plasmas that would expand into the surrounding dilute
background plasma, forming a shock front.  This is essentially an {\em
  explosive plasma shock}, just like that in an atmospheric explosion,
for which the downstream or background plasma is of similar or lower
temperature compared with the over-pressured exploding
plasma~\cite{sarri2011generation,sarri2010shock,dieckmann2010simulation}.
The physical picture of such an exploding plasma shock is well
established, as illustrated by the diagram in
Fig.~\ref{fig:diagram}(a).

Here we describe the contrasting case of an {\em ablative plasma
  shock} that has qualitatively different shock structures compared
with the exploding plasma shock previously studied. The canonical
setup for an ablative plasma shock is a cold/dense plasma in contact
with a hot/dilute background plasma, a situation that is also
ubiquitous in space, astrophysical, and laboratory plasmas.  The key
defining characteristic is the large ratio of hot and cold plasma
temperatures, $R \equiv T_h/T_c \gg 1.$ As we shall show, the primary
driver for the ablative plasma shock is the electron temperature
difference across the interface. Whether the two plasmas are in
pressure balance or not is less important.  This scenario of
electron-temperature-gradient or ETG-driven ablative plasma shock
occurs in, for example, the plasma thermal quench of a major
disruption in
tokamaks~\cite{zhang23cooling,zhang2023electron,li2023staged}, the
interfacial dynamics between central hot-spot plasmas and the
neighboring imploding liners or liner remnants in an inertial
confinement fusion (ICF)
capsule,~\cite{Lobatchev-Betti-PRL-2000,schiavi-atzeni-pop-2007,srinivasan-tang-pop-2014a,srinivasan-tang-epl-2014}
and the formation of structures in the galaxy
clusters~\cite{Fabian-ARAA-1994,Peterson-Fabian-PR-2006}. In such
cases, the hot-cold plasma interface is known to trigger a thermal
collapse of the nearly collisionless hot background plasma and an
ablative mix of the cold ions with the background hot
ions~\cite{zhang23cooling}. The latter becomes particularly
interesting if plasma ions on the two sides of the interface are of
different species~\cite{ji2008generating}, for example, in the high-Z
impurity pellet injection into fusion plasmas for tokamak disruption
mitigation, where efficient mixing of the ablated high-Z impurities
with the fusion plasma is critical for sufficiently uniform
radiations~\cite{mao2023rapid}. The ablative mix of metal into ICF
fuel is a similar physical case of practical importance.

The goal of this Letter is to give the physical picture and
characteristic properties of this new class of collisionless ablative
plasma shocks.  To establish the essential physics, we employ first
principles kinetic simulations with the VPIC code.~\cite{VPIC}
Although much of what is observed in the simulations can be
understood with theoretical analyses, we have deferred the far more
involved analytical analysis to a follow-up paper that allows space
for an adequate exposition.

Let's first introduce the characteristic structure of collisionless
ablative plasma shock and contrast it with that of the extensively
studied explosive plasma shock.  In an explosive plasma shock, the
shock propagates into the ambient plasma that has comparable or lower
electron temperature but much lower plasma density. The ambient plasma
is thus at the upstream with respect to the shock front, and the
over-pressured plasma is at the downstream.  For an ablative plasma
shock, the driver is the thermal energy flux from the hot and dilute
(background) plasma that is deposited into the cold and dense
plasma. The shock always propagates into the hot and dilute plasma,
independent of the pressure difference, as long as it is not too
extreme.  In the diagrams shown in Fig.~\ref{fig:diagram}, the
upstream is on the right side and the downstream is placed on the left
side, for both the ablative and explosive plasma shocks.  The
qualitative difference between the two shocks lies in the structure of
the electrostatic electric field. For the explosive plasma shocks, the
ambipolar potential is monotonically decreasing from the downstream to
the upstream across the shock region, producing an electric field that
traps the bulk electrons from the downstream over-pressured plasma.
Only the high-energy tail electrons from the downstream plasma can
escape this ambipolar potential trap. In contrast, all upstream
electrons can follow the increasing potential to accelerate through
the shock region into the downstream plasma.  As expected from a
monotonic electrostatic potential, the ions behave exactly the
opposite of the electrons, having a trapped population for upstream
ions while all downstream ions are accelerated by the ambipolar
electric field through the shock region.  Only high-energy tail ions
from the upstream can have a chance to go through the shock into the
downstream plasma. The collisionless explosive plasma shock has a
shock width on the order of the plasma Debye
length~\cite{taylor1970observation,romagnani2008observation,forslund1970formation},
which is delimited by two vertical black dashed lines in
Fig.~\ref{fig:diagram}(a).

In sharp contrast, the ablative plasma shock contains a more
complicated shock structure that has a Debye-scale shock region
sandwiched by much broader pre- and post-shock transition regions. The shock
region, which is delimited by two vertical dashed black lines in the
diagram of Fig.~\ref{fig:diagram}(b), is a surplus electron layer that
is negatively charged. There are two ion recession fronts in front of
and behind the shock front, marked by vertical dashed yellow lines in
the diagram of Fig.~\ref{fig:diagram}(b).  Between the ion recession
fronts and the shock layer, there are two surplus ion layers
sandwiching the electron layer of the shock front.  In contrast to the
electron or shock layer, which is non-neutral, the violation of
quasi-neutrality is weak in the two ion layers.  This is similar to
the transition layer that connects the quasi-neutral presheath to the
non-neutral plasma
sheath,~\cite{Li-etal-prl-2022,Franklin-Ockendon-JPP-1970} except that
we have two such transition layers smoothly transitioning the
non-neutral electron shock layer to the upstream and downstream
quasi-neutral plasmas.  Because the quasineutrality is weakly violated
in these two transition (ion) layers, their width can be much greater
than the Debye length and hence the shock layer.  The presence of the
two ion layers on both sides of the electron shock layer leads to a
non-monotonic electrostatic potential that reaches a minimum inside
the shock layer. The resulting ambipolar electric fields now reflect
thermal electrons in both upstream and downstream plasmas, as
illustrated in the diagram of Fig.~\ref{fig:diagram}(b).

The high energy tail electrons from both the upstream hot plasma and
the downstream cold plasma can escape their side of the electrostatic
trapping potential, which scales with the local hot/cold plasma
temperature. The tail electrons from the upstream provide the
conductive heating of the downstream plasma, mostly the electrons.
The tail electrons from the downstream, once they penetrate through
the shock front, ride the ambipolar potential in the upstream ion
layer to gain kinetic energy on the order of the hot plasma thermal
energy. This gives rise to a cold electron beam in the upstream
plasma.  Both the upstream hot ions and downstream cold ions are
accelerated by the non-monotonic electrostatic potential into the
shock front. The kinetic energy gain is proportional to the potential
drop between the respective recession front and the shock front, which
approximates the local initial hot/cold plasma temperature. As a
result, the upstream ion flow into the shock front can reach the hot
plasma sound speed, while the downstream ion flow into the shock is
much slower at the cold plasma sound speed.  Since there is a steep
potential well in the shock region on the side of the upstream, the
cold ions accelerated into the shock are promptly turned back, so the
shock front forms an effective barrier for the cold ions. The hot ions
have a kinetic energy of hot plasma thermal energy so they can easily
penetrate through the downstream ion layer. Their collisionless mixing
with the cold ions provides significant ion heating behind the
shock. This gives rise to the interesting phenomenon that
collisionless ablative shock can efficiently heat the ions but much
less so for the electrons behind the shock.

\begin{figure}[hbt]
\subfloat[]{\includegraphics[width=0.4\textwidth]{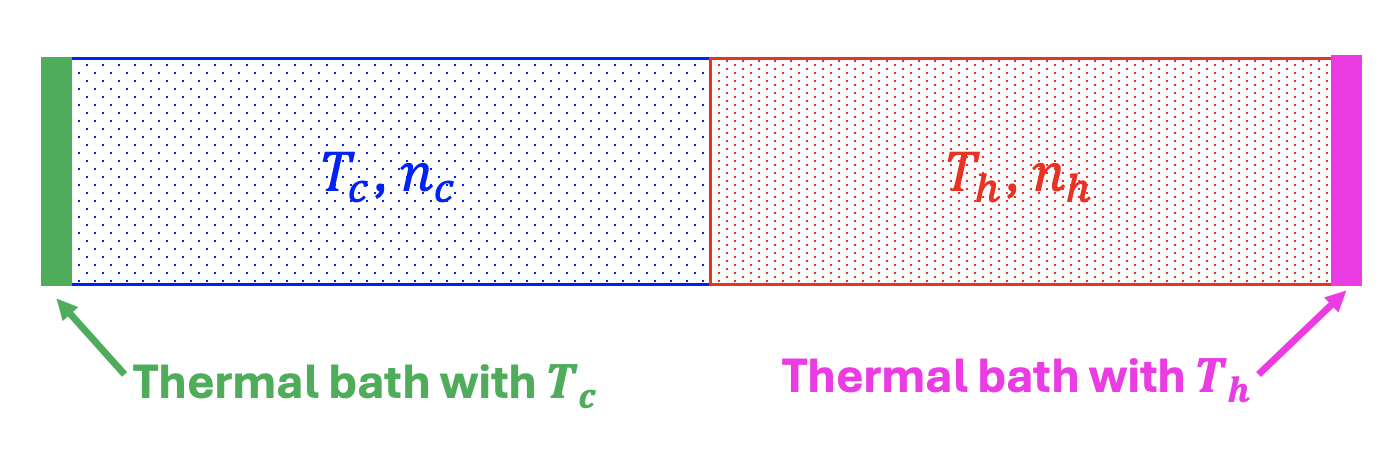}}
\hfill
\subfloat[]{\includegraphics[width=0.4\textwidth]{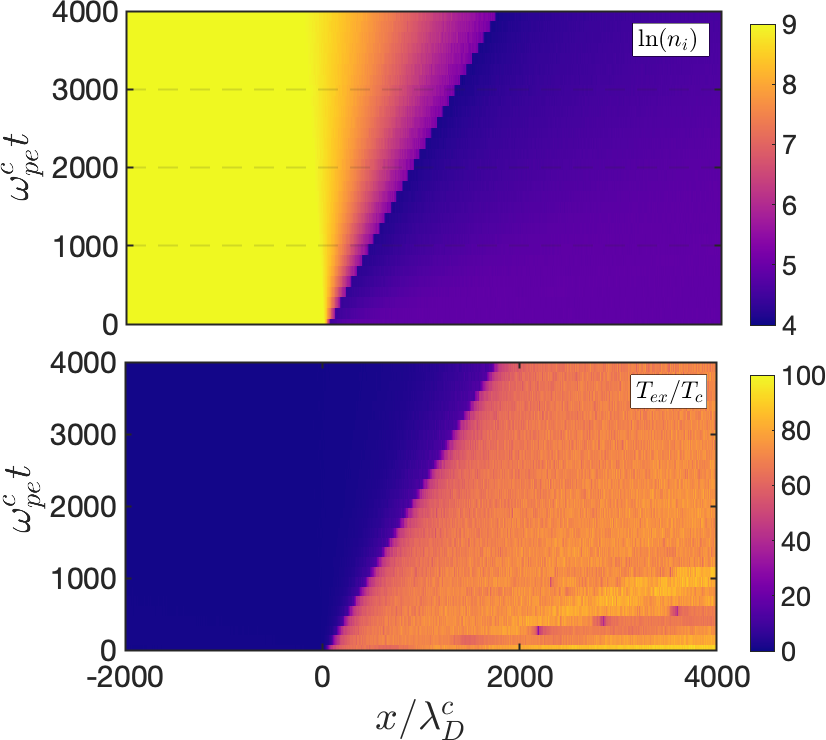}}
\caption{(a) Schematic view of VPIC simulation setup and (b)
  Spatial-temporal evolution of ion density and parallel electron
  temperature for $R=100$. Here $\omega_{pe}^c\propto \sqrt{n_c}$ and
  $\lambda_D^c\propto \sqrt{T_c/n_c}$ are, respectively, the cold
  plasma frequency and Debye length.}
\label{fig:simulation-setup}
\end{figure}

Having established the basic physical picture of the collisionless
ablative plasma shock, we will now provide additional quantitative
results from first principles kinetic simulations. As shown in
Fig.~\ref{fig:simulation-setup}(a), our VPIC simulations are
initialized by a slab plasma with two distinct sides of a cold/dense
plasma (density $n_c$ and temperature $T_c$) on the left and a
hot/dilute plasma (density $n_h$ and temperature $T_h$) on the right,
with a sharp interface initially.  At the two ends of the slab, the
boundary condition imposes the plasma contact with thermobaths of
temperature $T_c$ and $T_h,$ respectively.  The numerical
implementation is to compute the plasma particle flux leaving the
boundary and return them with a half-Maxwellian at temperature $T_c$
and $T_h,$ respectively.  The key dimensionless parameter in ablative
plasma shock is the temperature ratio $R\equiv T_h/T_c.$ The density
ratio $n_c/n_h$ is less important, and the shock physics has weak
dependence on whether the cold/dense plasma is over- or
under-pressured compared with the hot/dilute side, as long as it is
not too under-pressured with the pressure ratio
$p_c/p_h \gg\sqrt{1/R}.$ Unless
specifically noted, our presentation will be given using the
simulation results of the balanced pressure case ($p_c = p_h$).

\begin{figure}[hbt]
\includegraphics[width=0.35\textwidth]{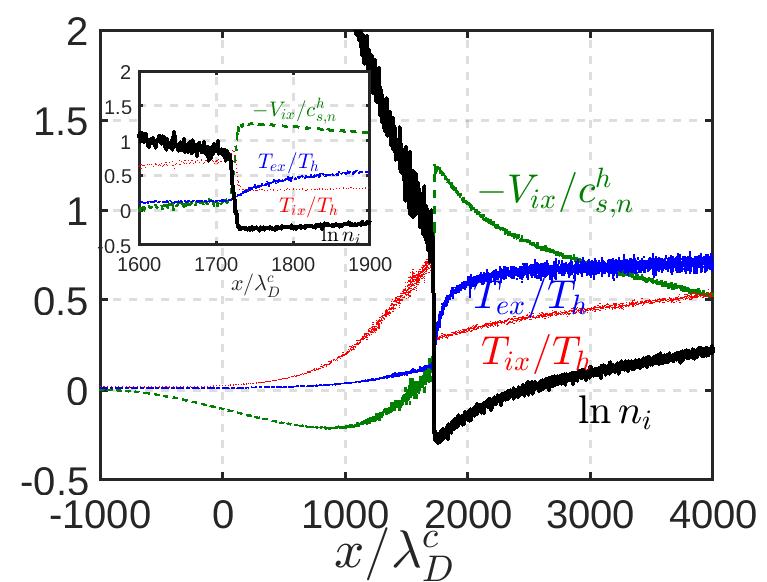}
\caption{Normalized plasma profiles ($\ln n_i$, $T_{e,ix}/T_h$, and
  $-V_{ix}/v_{th,i}^h$) corresponding to $\omega_{pe}^c t\approx 4000$
  in Fig.~\ref{fig:simulation-setup}(b), where the zoom-in plot shows
  the jumps near the shock front. Here the temperatures are normalized
  by $T_h$, while the flow is normalized by the nominal hot plasma
  sound speed $c_{s,n}^h\equiv \sqrt{T_h/m_i}$.}
\label{fig:shock-profile}
\end{figure}

\begin{figure}[hbt]
\centering
\subfloat[]{\includegraphics[width=0.4\textwidth]{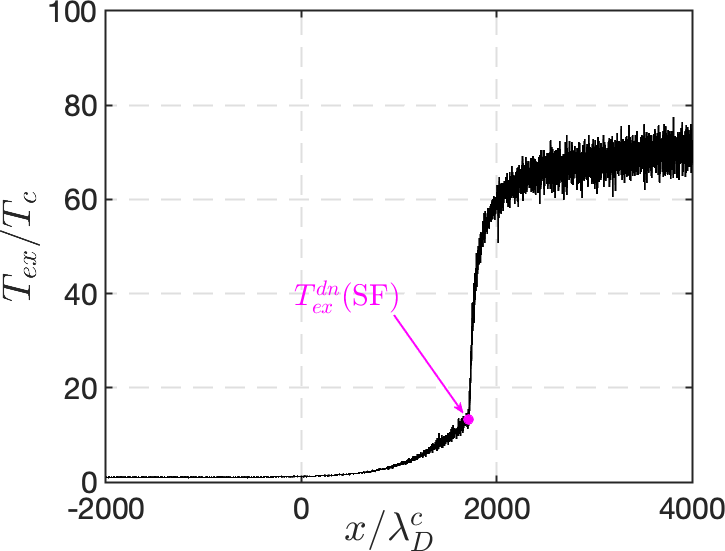}}
\hfill
\subfloat[]{\includegraphics[width=0.4\textwidth]{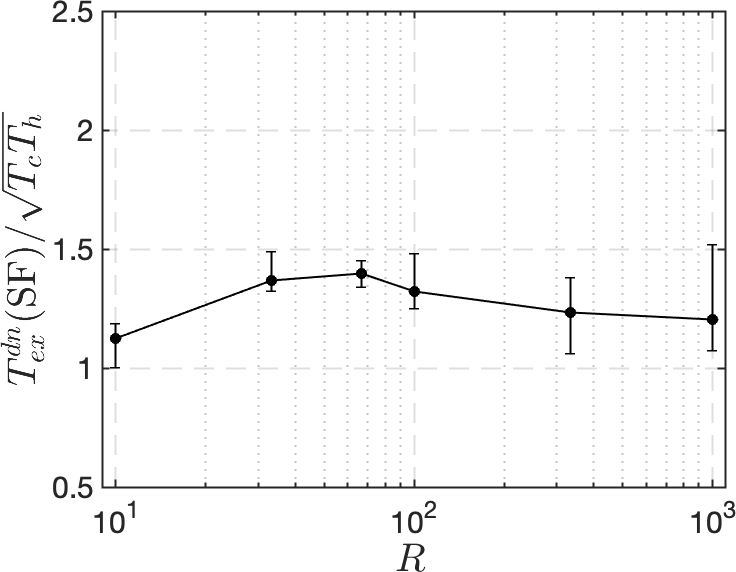}}
\caption{(a) Electron temperature profile across a collisionless
  ablative plasma shock for $R\equiv T_h/T_c=100$ and (b)
  $T_{ex}^{dn}({\rm SF})/\sqrt{T_cT_h}$ for different $R$.}
\label{fig:Te-shock}
\end{figure}

Both the upstream and downstream transition layers, which have surplus
ion charge densities, are of hot electron mean-free-paths
$\lambda_{h-h}$ and $\lambda_{h-c},$ with the subscript $h-h$ labeling
hot electron collisions in the upstream hot/dilute plasma, and $h-c$
the hot electron collisions with the downstream cold/dense plasma.
Collisionless ablative plasma shock occurs in the limit of
$\lambda_{h-h, h-c} \gg \lambda_D,$ so we will simply show the
collisionless VPIC simulation results to have a zoomed-in view of the
shock physics. From the plasma density and electron temperature
evolution in space and time shown in
Fig.~\ref{fig:simulation-setup}(b), one observes a shock, where both
plasma density and electron temperature have a jump, emanating from
the initial hot/cold interface and propagating into the hot/dilute
plasma. Focusing on a single time slice, we show the plasma behavior
around the shock in Fig.~\ref{fig:shock-profile}. The parallel
electron and ion temperatures ($T_{ex}$ and $T_{ix}$), ion density
($n_i$), and ion flow ($V_{ix}$) all have a jump at the shock
front. The upstream ion flow is of particular interest as it recovers
the so-called cooling flow popularized in plasma astrophysics, a
concept motivated by astronomical observations but suffers from the
lack of a plausible physics explanation from known plasma transport
physics.~\cite{Binney-Cowie-apj-1981,Fabian-etal-AAR-1991,Fabian-ARAA-1994,Peterson-Fabian-PR-2006}
There is also a downstream ion flow that is accelerated into the
shock. Since the upstream ions are passing through the shock with
little impedance from the downstream electric field, the downstream
ion flow into the shock is primarily made of cold ions being
accelerated by the downstream ambipolar electric field, see
Fig.~\ref{fig:diagram}(b).  The collisionless mixing of hot upstream
ions and the cold downstream ions behind the shock provides shock
heating of the ions to a temperature comparable to $T_h,$
Fig.~\ref{fig:shock-profile}. The upstream hot electrons suffer from
decompressional cooling associated with the accelerating cooling flow
into the shock front, and the conductive cooling from the divergence
of the electron heat conduction flux, which also has a contribution
from the cold electron beam that leaks through the shock front from
the downstream plasma.  The conductive heating of the downstream
electrons, which is also a result of collisionless mixing, is far less
effective compared with the shock heating of the ions. The rise in
electron temperature just behind the shock takes the geometric mean of
$T_c$ and $T_h.$ Specifically, Fig.~\ref{fig:Te-shock}(a) shows the
$T_{ex}$ profile in the neighborhood of the shock, with
$T_{ex}^{dn}(\rm{SF})$ labeling the downstream parallel electron
temperature at the shock front, while Fig. 4(b) demonstrates the
simulation results of
\begin{align}
T_{ex}^{dn}(\rm{SF}) \approx \sqrt{T_c T_h},
\end{align}
for a wide range of $R\equiv T_h/T_c$ values.

\begin{figure}[hbt]
\centering
\subfloat[]{\includegraphics[width=0.4\textwidth]{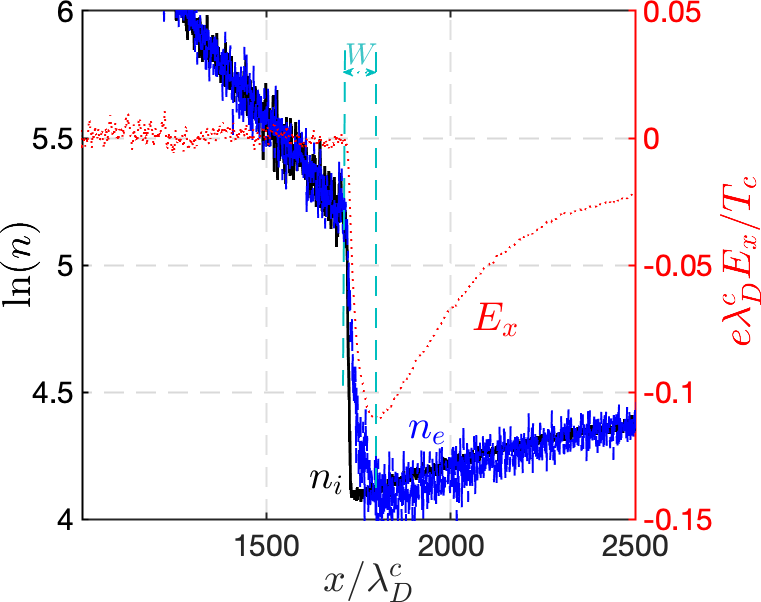}}
\hfill
\subfloat[]{\includegraphics[width=0.35\textwidth]{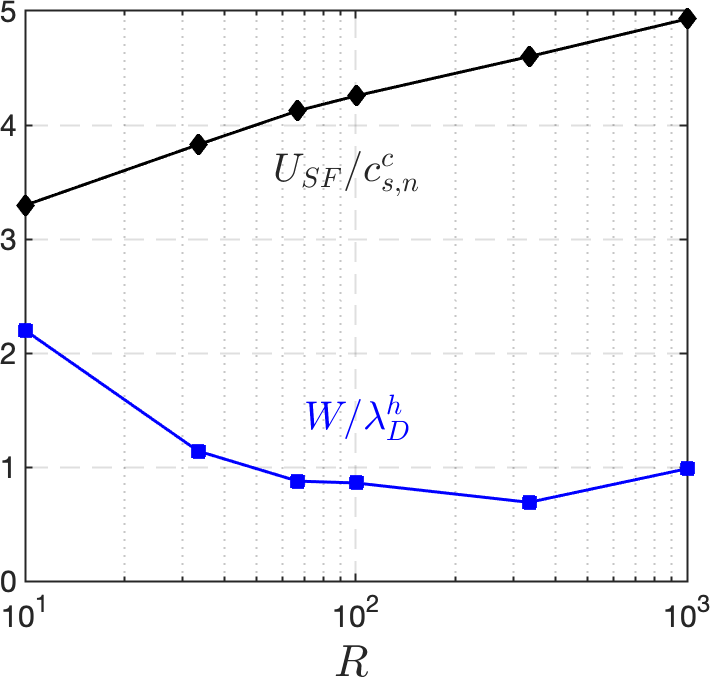}}
\caption{(a) Ion and electron density (left y-axis) as well as the
  electric field (right y-axis) for $R\equiv T_h/T_c=100$, where two
  vertical cyan lines separate the quasi-neutral plasmas (the ion
  layers) away from and the non-neutral plasma (the electron layer) at
  the shock front. (b) The shock front width $W$ and speed $U_{SF}$
  for different $R$.}
\label{fig:shock-width}
\end{figure}

It is important to note that in the normal shock theory where thermal
transport is ignored, the shock front speed $U_{\textrm{SF}}$ would
have scaled with the local sound speed set by the local electron
temperature $T_{ex}^{dn}(\textrm{SF}),$ so the shock speed would have
a strong $R$ scaling,
\begin{align}
U_{\textrm{SF}} \propto \sqrt{T_{ex}^{dn}(\textrm{SF})/m_i}
\sim R^{1/4} \sqrt{T_c/m_i} = R^{1/4} c_{s,n}^c.
\end{align}
When the thermal flux is accounted for, the shock speed is known to be
able to decouple from the local electron temperature
$T_{ex}^{dn}.$~\cite{kuznetsov2018parallel} In the case of a
collisionless ablative plasma shock, the transport modification takes
the extreme form that the shock front speed $U_{\textrm{SF}}$
completely loses the $R^{1/4}$ dependence and simply scales with the
cold plasma sound speed $c_{s,n}^c=\sqrt{T_c/m_i},$
\begin{align}
U_{\textrm{SF}} \sim c_{s,n}^c.
\end{align}
Fig.~\ref{fig:shock-width} plots the shock front speed from the
simulations as a function of $R\equiv T_h/T_c.$ It is remarkable that
a rough fitting of the simulation data shows an $R$ correction
entering as a logarithmic term,
\begin{align}
U_{\textrm{SF}} \approx \left(2.6 + 0.35\ln R\right) c_{s,n}^c,
\end{align}
in the range of $R\in \left(10, 10^3\right).$

The width of the collisionless ablative plasma shock is found in the
simulations to correspond to a well-defined (surplus) electron charge
layer, Fig.~\ref{fig:shock-width}(a).  Contrary to the shock front
speed, the width of this electron layer is set by the upstream plasma
conditions,
\begin{align}
W \sim \lambda_{D}^h.
\end{align}
In Fig.~\ref{fig:shock-width}(b), we show the simulation data that has
the shock width, scaled by the upstream hot plasma Debye length,
nearly independent of $R\equiv T_h/T_c,$ in the range of $R\in
\left(10, 10^3\right).$

In conclusion, we have uncovered the shock structure of a class of
collisionless ablative plasma shock, which can occur in an array of
space, astrophysical, and laboratory plasmas where a cold/dense plasma
is in contact with a hot/dilute ambient plasma. First-principle
kinetic simulations have revealed an electrostatic potential well
across the shock front. This is generated by a negatively charged
electron shock layer sandwiched by two weakly positively charged ion
transition layers in both upstream and downstream plasmas. The
potential well has its bottom in the shock layer, reflecting bulk
electrons from both sides but accelerating both downstream cold ions
and upstream hot ions into the shock front. The collisionless mixing
of upstream hot ions and downstream cold ions produces efficient shock
heating of the ions behind the shock.  In contrast, the electron
heating behind the shock is much milder with a temperature about the
geometric mean of the hot and cold electrons. The width of a
collisionless ablative plasma shock scales with the upstream hot
plasma Debye length, while the shock speed scales with the sound speed
of the downstream cold plasma. Both the qualitative picture and
quantitative predictions on the collisionless ablative plasma shock
provide intriguing targets for experimental investigations and further
analysis.

\textbf{Acknowledgment} We thank the U.S. Department of Energy Office
of Fusion Energy Sciences through the Base Fusion Theory Program at
Los Alamos National Laboratory (LANL) under contract
No. 89233218CNA000001. This research used resources of the National
Energy Research Scientific Computing Center, a DOE Office of Science
User Facility supported by the Office of Science of the
U.S. Department of Energy under Contract No. DE-AC02-05CH11231 using
NERSC award FES-ERCAP0032298 and LANL Institutional Computing Program,
which is supported by the U.S. Department of Energy National Nuclear
Security Administration under Contract No. 89233218CNA000001.

%\bibliography{reference}% Produces the bibliography via BibTeX.


\end{document}